\documentclass[prl,aps,nopacs,superscriptaddress,nofootinbib]{revtex4}
\usepackage{graphicx} 
\usepackage{dcolumn}  
\usepackage{bm}       
\usepackage{amsmath}
\usepackage{epsfig}
\begin{document}

\title{Plasmonic absorption properties of bimetallic metamaterials}
\author{Evangelos~Atmatzakis}
\affiliation{Optoelectronics Research Centre and Centre for Photonic Metamaterials, University of Southampton, Southampton SO17 1BJ, United Kingdom}
\author{Nikitas~Papasimakis}
\affiliation{Optoelectronics Research Centre and Centre for Photonic Metamaterials, University of Southampton, Southampton SO17 1BJ, United Kingdom}
\author{Nikolay~I.~Zheludev}
\email{niz@orc.soton.ac.uk}
\affiliation{Optoelectronics Research Centre and Centre for Photonic Metamaterials, University of Southampton, Southampton SO17 1BJ, United Kingdom}
\affiliation{The Photonics Institute and Centre for Disruptive Photonic Technologies, Nanyang Technological University, Singapore 637371}
\date{\today}

\begin{abstract}
We demonstrate polarization controlled absorption in plasmonic bimetallic metamaterials. We fabricate and experimentally characterize Au/Ni ring resonator arrays, where by varying the wavelength and polarization of the incident wave, local electromagnetic fields and dissipation can be suppressed or enhanced in the Au and Ni areas of the rings.
\end{abstract}

\maketitle

\section{Introduction}
Plasmonic metamaterials exhibit extraordinary properties that often require a resonant dispersion. However, intrinsic losses in plasmonic systems usually damp the optical response and result in weaker resonances. On the other hand, there are applications which require strong optical absorption and maximizing the dissipation losses is actually desirable \cite{BHN15}. A few examples include the use of plasmonic metamaterials in order to enhance photoemission \cite{HS68,SB81,Zhukovsky2013}, form water vapour \cite{Halas}, generate electricity \cite{C14,plasmoelectric} and thermoelectrically driven magnetic pulses \cite{thermo,Guillaumme}. In principle, the increased absorption cross-section that nanostructures offer \cite{B83} is an effective mechanism to transfer energy from light to matter. Since absorption depends on geometrical and material properties, realizing a dynamically controlled system can be challenging. For example, it has been demonstrated that a thin-film perfect absorber with tunable absorption can be realized by taking advantage of interference effects between two incident light beams \cite{Xu14,Zhang2012}. In this work we follow a different approach and propose the use of a composite resonator, comprising materials with significantly different optical conductivities, which enables the tailoring of the system's absorption by manipulating the optical mode and promoting current flow in either part.  
\begin{figure}[ht]
	\includegraphics[width=8cm]{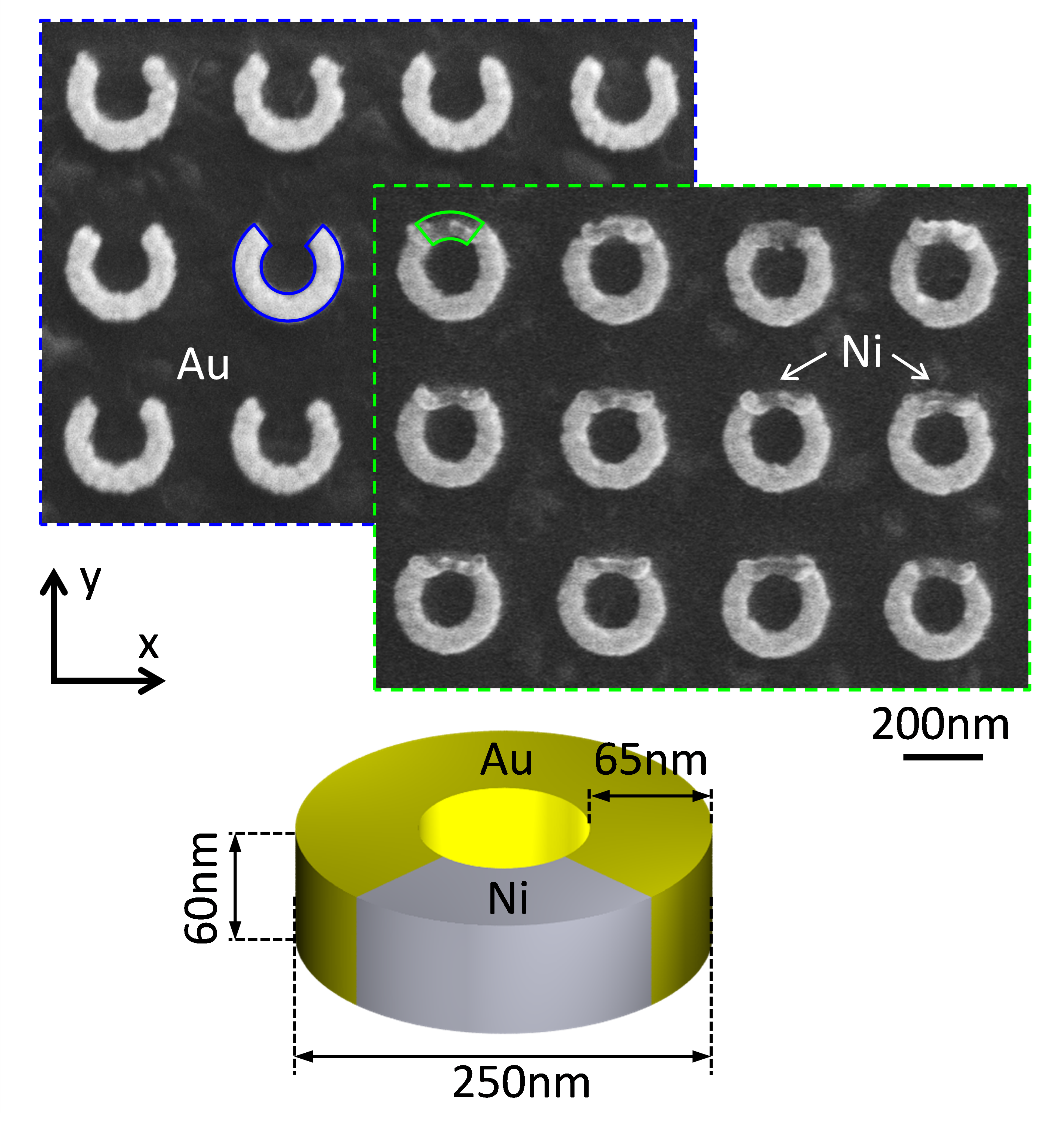}
	\caption{{\bf Scanning Electron Microscope (SEM) images of the bimetallic ring resonators}. An example of bimetallic ring resonators consisting of 3/4 Au and 1/4 Ni. The left image shows the incomplete ring formed by the Au part and on the right is an image of the final sample. The rendering on the bottom is a graphical representation of the ring that has been simulated numerically.} 
	\label{Fig1}
\end{figure}
We introduce a plasmonic metamaterial consisting of bimetallic Au/Ni ring resonators, and study its absorption properties. We demonstrate strong plasmonic resonances which depend on the composition of their constituent metals, as well as the wavelength and polarization of the incident wave. Also, due to the non-uniformity of the unit-cell, the two ring sections exhibit spectrally separated and polarization sensitive optical absorption. 

\begin{figure*}[ht]
	\includegraphics[width=16cm]{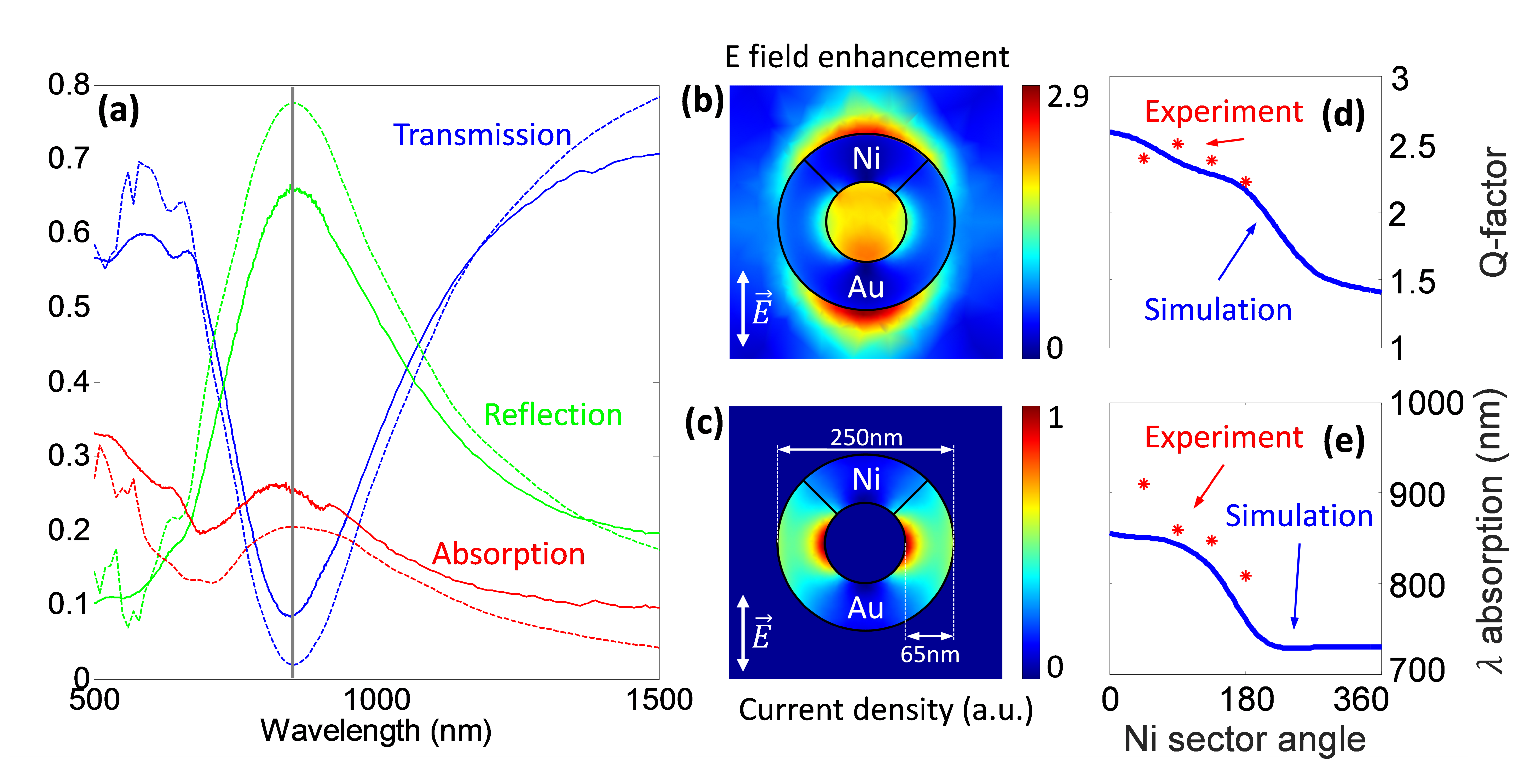}
	\caption{{\bf Optical properties of bimetallic ring resonators}. {\bf (a)} Transmission, reflection and absorption spectra of a resonator with a $90^{\circ}$ Ni sector and $270^{\circ}$ Au sector. Solid lines represent fabricated samples and dashed lines the numerical simulations. {\bf (b)} Electric field enhancement map on the vicinity of ring. {\bf (c)} Current density distribution along the ring. {\bf (d)} Numerical (blue) and experimental (red) measurement of Q-factor for 4 different bimetallic resonators where the Ni sectors extend in $45^{\circ}$, $90^{\circ}$, $135^{\circ}$ and $180^{\circ}$. {\bf (e)} The spectral position of the corresponding peaks of absorption in simulation and experiment.}
	\label{Fig2}
\end{figure*}

\section{Design and fabrication}
We study arrays of bimetallic ring resonators with $185\,$nm mean diameter ($65\,$nm of linewidth) consisting of Au and Ni (see Fig.\ \ref{Fig1}). The angle of the Ni sector controls the composition of the two metals in the ring. Four samples have been fabricated with Ni sectors that span over $45^{\circ}$, $90^{\circ}$, $135^{\circ}$ and $180^{\circ}$. The samples were fabricated by a two-step electron-beam lithography method on a glass substrate. Thin metal films, with thickness of $60\,$nm were thermally evaporated on the substrate and lifted-off to reveal each sector of the ring. To facilitate a faster lift-off process, a $50\,$nm thick layer of co-polymer was deposited prior to $200\,$nm of PMMA. As it can be seen in Fig.\ \ref{Fig1}, during the first step Au split-rings are formed (blue outline) and subsequently the Ni part (green outline) is placed to complete the ring, ensuring that the two metal parts form solid junctions. In order to account for fabrication inaccuracies and alignment limitations the two sectors were designed to overlap over a small area around the junctions. A good contact between the metal sectors is essential in order for the system to respond according to design, as any gap caused by fabrication error strongly affects the position and strength of plasmonic resonances. The surface roughness of the fabricated samples ranges between $5$ and $20\,$ nm and is present mainly at the nickel sections. As Ni tends to form large grains, small deformations usually develop on its surface, resulting in high values of roughness. However, these deformations are limited only to the direction normal to the metamaterial plane as all other surfaces are strictly defined during the lithography process or by the substrate. 

\section{Results}
The fabricated samples were characterized optically using a commercial microspectrometer system and imaged under a scanning electron microscope. Numerical simulations of the metamaterial's plasmonic response were carried out using Comsol Multiphysics, a commercial finite-element solver.

Upon excitation of the ring with light polarized along the axis of symmetry, a strong plasmonic resonance appears centered around $850\,$nm (Fig.\ \ref{Fig2}a), which corresponds to the lowest-order dipole mode of the ring. The latter is inferred by examining the field maps of Fig.\ \ref{Fig2}b, where two areas of electric field enhancement can be identified across the ring. Accordingly, the current distribution is at a maximum in the region between the two electric field enhancement areas (see Fig.\ \ref{Fig2}c). The composition of the ring resonators strongly affects the response of the system, as the two metals have different optical properties. Nickel is less conductive and exhibits higher losses due to Joule heating in comparison to Au. Hence, increasing the size of the Ni sector leads to damping of the resonance and a decrease of the Q-factor (Fig.\ \ref{Fig2}d). At the same time, the position of the resonance peak blue-shifts due to changes in the effective permittivity of the ring (Fig.\ \ref{Fig2}e).  
Exploiting the difference in conductivities between the two metals that constitute the resonator, we can realize a system whose level and spatial distribution of optical absorption can be dynamically controlled. In the bimetallic resonators, optical absorption is driven by dissipation losses which depend on both the respective size and current density in each metal. The latter is defined by the plasmonic mode that is excited in the ring and is controlled by the polarization and wavelength of incident light. Based on that, we define three methods of controlling absorption.
\begin{figure}[ht]
	\includegraphics[width=8cm]{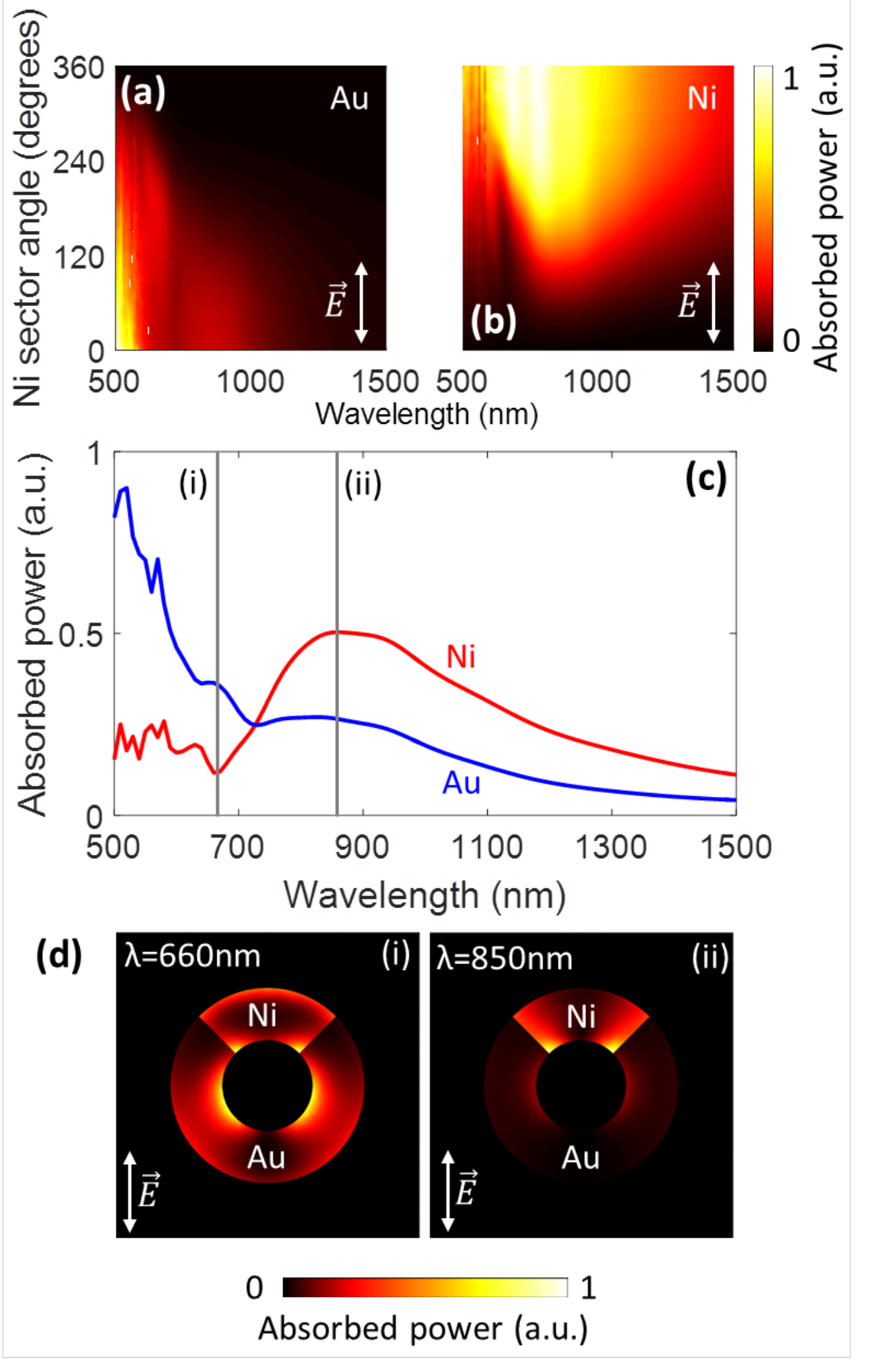}
	\caption{{\bf Material and wavelength selective absorption.} Spectrum of absorption in the Au {\bf (a)} and Ni {\bf (b)} part in relation to the angle of the Ni sector. {\bf (c)} The absorption spectrum of the two different parts for the case of $90^{\circ}$ Ni sector. {\bf (d)} Distribution of absorbed power in arbitrary units across the ring in two characteristic spectral positions marked with grey lines in (c).}
	\label{Fig3}
\end{figure}

{\bf Material selective absorption.} Altering the size of the two metal sectors, is an effective way of changing the dissipation losses in the system. In order to gain an insight on how each metal contributes to the total absorption for different compositions, we look at the Au and Ni sectors individually (see Fig.\ \ref{Fig3}a,b). At the plasmonic resonance ($850\,$nm), absorption occurs predominantly in Au for smaller Ni sections, while for larger ones the total absorption increases and takes places mainly in the nickel part of the ring. This behaviour is attributed to the higher losses of Ni, which rapidly dominate the absorption of the bimetallic system as the size of the Ni section increases. Additional absorption peaks can also be identified in both Au and Ni at the blue end of the spectrum linked to lattice and higher-order plasmonic modes. These modes engage both metals as they are generally less confined and spread along the ring.

{\bf Wavelength selective absorption.} As absorption depends on the current density mode that is excited on the resonator, we can engage different absorption regimes by shifting the maxima of the current density towards the more or less conductive regions of the resonator. In Fig.\ \ref{Fig3}c, we show that varying the excitation wavelength of a ring with a $90^{\circ}$ Ni sector enables the shift of absorption from Au (at shorter wavelengths) to Ni (at $\sim 850\,$nm). This is further illustrated in Fig.\ \ref{Fig3}d, where the spatial distribution of absorbed power is presented. At $660\,$nm, absorption occurs mainly in the Au part of the ring, while at $850\,$nm the situation is reversed. This shift of currents away from the Ni section for decreasing wavelength, can be linked to the transitioning from the fundamental mode to a higher order ones. 

\begin{figure*}[ht!]
\includegraphics[width=16cm]{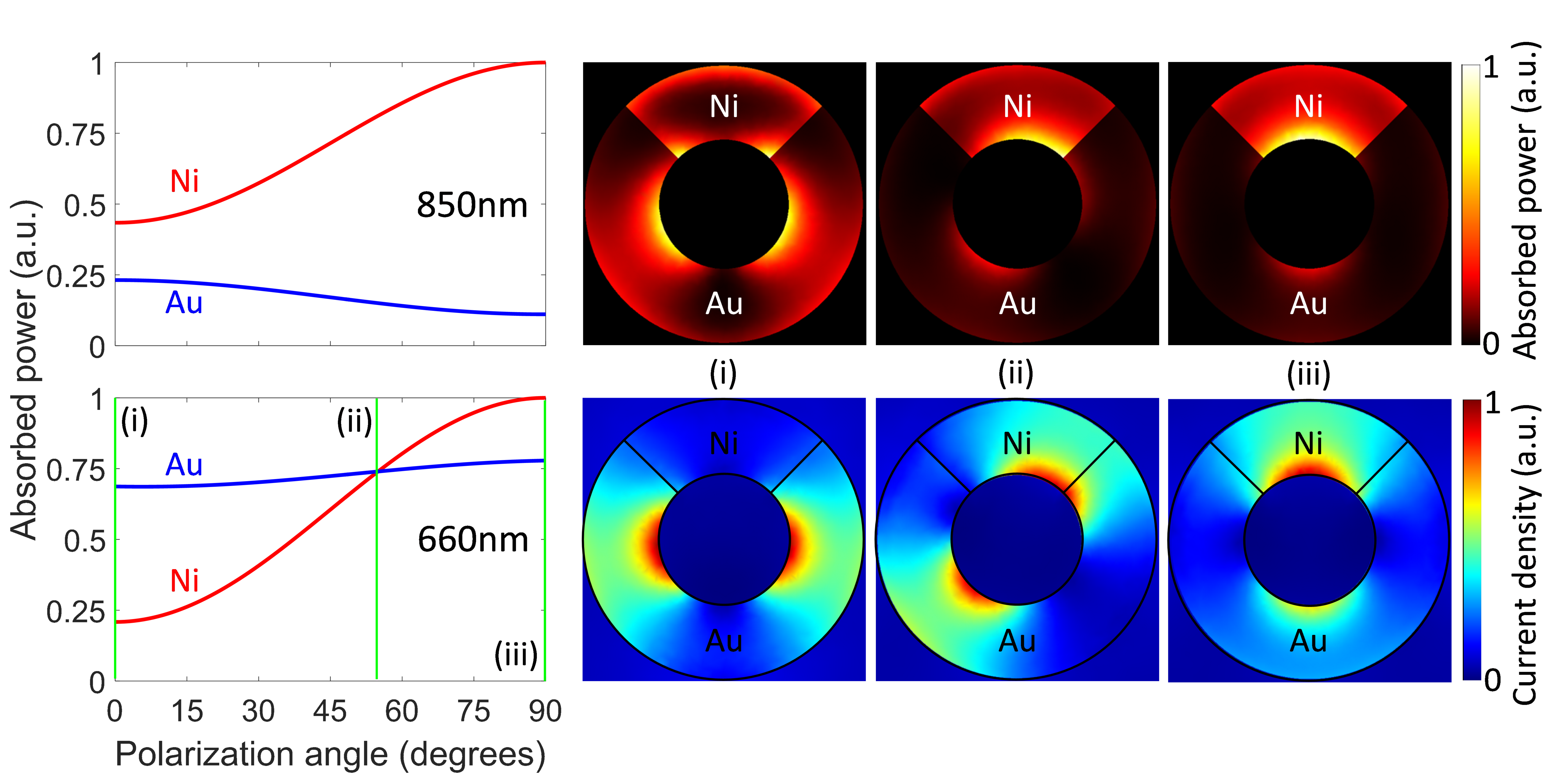}
\caption{{\bf Polarization selective absorption.} Absorption in Au (blue) and Ni (red) part of a resonator with $90^{\circ}$ Ni sector at different polarization angles of the incident beam. The top graph corresponds to excitation at the resonance of the ring (850 nm) and the bottom at the point where absorption in Ni is lowest (660 nm). Distribution of the absorbed power on the ring and the corresponding current density for different polarization angles marked with the green lines.}
\label{Fig4}
\end{figure*}

{\bf Polarization selective absorption.} The current configuration of the bimetallic rings at resonance can also be controlled by simply rotating the polarization azimuth of the incident beam, which allows us to steer the current density towards or away from the lossy Ni section. This mechanism of polarization selective absorption becomes more clear when examining a specific case of a bimetallic ring with a Ni sector of $90^{\circ}$. In Fig.\ \ref{Fig4} we present the absorption in the Au and Ni sectors as a function of the incident wave's polarization. When the excitation beam is linearly polarized and parallel to y-axis (see Fig.\ \ref{Fig1}), denoted here as $0^{\circ}$, the currents (and hence absorbed power) in Ni are at a minimum (Fig.\ \ref{Fig4}(i)). Rotation of the polarization angle away from this position is accompanied by a shift of the current distribution and the corresponding absorption "hot-spots", engaging larger portions of the Ni section. This transitioning from polarization angles between $0^{\circ}$ and $90^{\circ}$ can be seen in Fig.\ \ref{Fig4}(i)-(iii). The final position, marks the polarization at which the current density and absorption that are induced in the Ni sector are at maximum. This method of manipulating the plasmonic mode through the polarization of incident beam enables control not only over the level of absorption but also its spatial distribution. 

\section{Conclusion}
In summary, we have shown that bimetallic ring resonator arrays excited at resonance exhibit strong absorption, which can be tuned by varying the ring composition, as well as polarization and wavelength of the incident light. Our results offer an effective way to control field enhancement and consequently dissipation losses at particular areas of the ring resonators, which can be used in applications where site-specific heating is essential. Finally, bimetallic hybrids exhibit a variety of interesting properties like differential thermal expansion or the ability to generate thermoelectric currents, that can be used, for example, in the generation of THz magnetic pulses \cite{thermo,Guillaumme}.  

\section{Acknowledgements}
The authors would like to thank Anagnostis Tsiatmas and F. Javier Garcia de Abajo for numerous discussions. The authors acknowledge the support of the MOE Singapore (grant MOE2011-T3-1-005), the UK’s Engineering and Physical Sciences Research Council (grants EP/G060363/1), and the Leverhulme Trust.


\begin{thebibliography}{10}
	
	\bibitem{BHN15}
	Mark~L. Brongersma, Naomi~J. Halas, and Peter Nordlander.
	\newblock {Plasmon-induced hot carrier science and technology}.
	\newblock {\em Nature Nanotechnology}, 10(1):25--34, 2015.
	
	\bibitem{HS68}
	J.~Hofmann and W.~Steinmann.
	\newblock Plasma resonance in the photoemission of silver.
	\newblock {\em physica status solidi (b)}, 30(1):K53--K56, 1968.
	
	\bibitem{SB81}
	J.~E. Sipe and J.~Becher.
	\newblock Surface-plasmon-assisted photoemission.
	\newblock {\em J. Opt. Soc. Am.}, 71(10):1286--1288, Oct 1981.
	
	\bibitem{Zhukovsky2013}
	S.~V. Zhukovsky, V.~E. Babicheva, A.~V. Uskov, I.~E. Protsenko, and A.~V.
	Lavrinenko.
	\newblock Enhanced electron photoemission by collective lattice resonances in
	plasmonic nanoparticle-array photodetectors and solar cells.
	\newblock {\em Plasmonics}, 9(2):283--289, 2013.
	
	\bibitem{Halas}
	O.~Neumann, A.~S. Urban, J.~Day, S.~Lal, P.~Nordlander, and N.~J. Halas.
	\newblock Solar vapor generation enabled by nanoparticles.
	\newblock {\em ACS Nano}, 7(1):42--49, 2013.
	
	\bibitem{C14}
	C{\'{e}}sar Clavero.
	\newblock {Plasmon-induced hot-electron generation at nanoparticle/metal-oxide
		interfaces for photovoltaic and photocatalytic devices}.
	\newblock {\em Nature Photonics}, 8(2):95--103, 2014.
	
	\bibitem{plasmoelectric}
	M.~T. Sheldon, J.~van~de Groep, A.~M. Brown, A.~Polman, and H.~A. Atwater.
	\newblock Plasmoelectric potentials in metal nanostructures.
	\newblock {\em Science}, 346(6211):828--832, 2014.
	
	\bibitem{thermo}
	A.~Tsiatmas, E.~Atmatzakis, N.~Papasimakis, V.~A. Fedotov, B.~Luk'yanchuk,
	N.~I. Zheludev, and F.~J. García~de Abajo.
	\newblock Optical generation of intense ultrashort magnetic pulses at the
	nanoscale.
	\newblock {\em New Journal of Physics}, 15(11):113035, 2013.
	
	\bibitem{Guillaumme}
	G.~Vienne, X.~Chen, Y.~S. Teh, Y.~J. Ng, N.~O. Chia, and C.~P. Ooi.
	\newblock Novel layout of a bi-metallic nanoring for magnetic field pulse
	generation from light.
	\newblock {\em New Journal of Physics}, 17(1):013049, 2015.
	
	\bibitem{B83}
	C.~F. {Bohren}.
	\newblock How can a particle absorb more than the light incident on it?
	\newblock {\em American Journal of Physics}, 51:323--327, 1983.
	
	\bibitem{Xu14}
	X.~Fang, M.~L. Tseng, J.-Y. Ou, K.~F. MacDonald, D.~P. Tsai, and N.~I.
	Zheludev.
	\newblock Ultrafast all-optical switching via coherent modulation of
	metamaterial absorption.
	\newblock {\em Applied Physics Letters}, 104(14):141102, April 2014.
	
	\bibitem{Zhang2012}
	J.~Zhang, K.~F. MacDonald, and N.~I. Zheludev.
	\newblock {Controlling light-with-light without nonlinearity}.
	\newblock {\em Light: Science {\&} Applications}, 1(7):e18, 2012.
	
\end{thebibliography}
\end{document}